\documentclass[11pt, a4paper]{article}
\usepackage{epsfig,enumerate,amsmath,amsfonts,amssymb}
\usepackage{epsf}
\usepackage{subfigure,psfrag,slashed,caption}
\usepackage{graphicx}
\usepackage{dcolumn}
\usepackage{bm}
\usepackage{capt-of}
\begin{document}
\title{Strong Decay Widths and Coupling Constant of Recent Charm Meson States}
\author{Meenakshi Batra, Alka Upadhayay \\\small{\it School of Physics and Material Science}\\ \small{\it Thapar University, Patiala-147004, India}\\\small{E-mail: mbatra310@gmail.com, alka@thapar.edu}}
\maketitle
\begin{abstract}
Open charm hadrons with strange and non-strange mesons have been
discovered in recent years. We study the spectra of several newly
observed resonances by different collaborations like BaBar
\cite{Sanchez}and LHCb \cite{Aaij} etc. Using an effective
Lagrangian approach based on heavy quark symmetry and chiral
dynamics, we explore the strong decay widths and branching ratios of
various resonances and suggest their $J^{p}$ values. We try to fit
the experimental data to find the coupling constants involved in the
strong decays through pseudoscalar mesons. The present work also
discusses about the possible spin-parity assignments of recently
observed states by LHCb collaboration. The tentative assignment of
newly discovered state $D_{J}^{\ast}(3000)$ can be natural parity
states $(0^{-},1^{+},2^{-},3^{+}....)$ while $D_{J}(3000)$ can be
identified with unnatural parity states like
$(0^{+},1^{-},2^{+},3^{-}....)$. Therefore, the missing doublets
2S,1D,1F,2P and 3S can be thought of filled up with these states. We
study the two-body strong decay widths and branching ratios of
missing doublets and plot branching ratios vs mass of decaying
particle. These plots are used to analyze all assignments to
$D_{J}(3000)$ deeply and various possibilities for $J^{P}$ values.
\vspace{0.2cm}
\end{abstract}
\section{Introduction}
The hadrons containing a single heavy quark can be analyzed in a
framework which is formulated for $N_{f}$ heavy quarks with mass
$m_{Q}>>\Lambda_{QCD}$ using heavy quark effective theory
\cite{Neubert}. This theory assumes heavy quark to act as static
color source and its spin is $s_{Q}$ which can be thought of coupled
to spin of light degrees of freedom $s_{l}$. Secondly, heavy quark
flavor symmetry leads to interaction of heavy quark with light
degrees of freedom through exchange of soft gluons only. The heavy
quark spin-flavor symmetry can be exploited further to explore
several hadronic properties. The motivation for present work arises
due to recently observed charm and bottom meson states by
experimental collaborations like BaBar, LHCb and CDF
\cite{Aaltonen}. The heavy meson spectrum is one of the recent
interest to place for the various particles at different resonances
and energies. Recently, some excited charm meson states were
observed which are D(2550), D(2600), D(2750) and D(2760)in the decay
channels $D^{0}(2550)\rightarrow
D^{\ast+}\pi^{-}$,$D^{0}(2600)\rightarrow
D^{\ast+}\pi^{-},D^{+}\pi^{-}$,$D^{0}(2750)\rightarrow
D^{+}\pi^{-}$,$D^{+}(2600)\rightarrow D^{0}\pi^{+}$ and
$D^{+}(2760)\rightarrow D^{0}\pi^{+}$ in the inclusive
$e^{+}e^{-}\rightarrow c\bar{c}$ interactions by BaBar Collaboration
\cite{Sanchez}. The most suitable spin-parity assignments for
D(2750),D(2760)is $(2^{-},3^{-})$ or 1D and for D(2550), D(2600)is
$(0^{-},1^{-})$ i.e 2S state respectively. LHCb collaboration
\cite{Aaij} observed some new resonances in addition to above i.e.
$D_{J}(3000)^{+,0}$ and $D_{J}^{\ast}(3000)^{0}$ around 3GeV in
association with $D_{2}^{\ast}(2460)^{0}$ and
$D_{J}^{\ast}(2760)^{0}$ and exists in the $D^{+}\pi^{-}$ invariant
mass spectrum. The states,
$D_{J}^{\ast}(3000)^{+}$,$D_{2}^{*}(2460)^{+}$ and
$D_{J}^{\ast}(2760)^{+}$ were observed in $D_{0}\pi^{+}$ spectrum
whereas the states measured in $D^{*+}\pi^{-}$ spectrum were
$D_{1}(2420)^{0}$,$D_{2}^{\ast}(2460)^{0}$,
$D_{J}^{\ast}(2760)^{0}$,$D_{J}(2580)^{0}$, $D_{J}(2740)^{0}$ and
$D_{J}(3000)^{0}$ respectively. A table for the properties of
recently observed states has been shown below. Similar is the case
in beauty sector. Very recently CDF Collaboration \cite{Aaltonen}
has found evidence for a new resonance B(5970) simultaneously in
$B^{0}\pi^{+}$ and $B^{+}\pi^{-}$ mass distributions with a
significance of 4.4$\sigma$ standard deviations and further reported
the first study of resonances with orbitally excited $B^{+}$ mesons
and updated measurement of orbitally excited $B^{0}$ and $B_{s}^{0}$
mesons. The branching ratio for $B_{s2}^{\ast 0}\rightarrow
B^{\ast+}K^{-}$ decays is also measured. The masses of new B(5970)
measured resonances are $5978\pm 5(stat)\pm 12(syst)MeV/c^{2}$ for
neutral state and $5961\pm5(stat)\pm 12(syst)MeV/c^{2}$ for charged
asymmetry into $B\pi$ states. This state may be proposed as
belonging to radially excited bottom meson family. Therefore, in
past decades, we faced several ground as well as excited states of
charm meson family such as discovery of $D_{sJ}(2460)$,
$D_{sJ}(2632)$ and $D_{sJ}(3040)$ etc. In bottom meson family we
witness some new states such as B(5279), $B^{\ast}(5325)$ for n=1
family in $(0^{-},1^{-})$ doublet. In the infinite heavy quark mass
limit, a heavy light system $Q\overline{q}$ can be classified into
doublets depending upon their quantum numbers. A heavy hadronic
system containing heavy quark with spin quantum number $S_{Q}$ and
light degrees of freedom $s_{l}$ that include light quark and gluons
interacting through quark-antiquark pairs. It should have the
quantum number of light quark that is $S_{l}$ in order to have total
conserved quantum number J where $J=S_{Q}+S_{l}$. Defining J as
$J^{2}=j(j+1)$ and $S_{Q}^{2}=(s_{Q})(s_{Q}+1)$ and
$S_{l}^{2}=(s_{l})(s_{l}+1)$,the total spin
$j\pm=s_{l}\pm\frac{1}{2}$ can be obtained by combining the spin of
heavy quark spin $\frac{1}{2}$ with spin of light degrees of
freedom. The heavy mesonic system form a degenerate doublet of
ground state with $J=\pm1$ and negative parity denoted as D and
$D^{\ast}$ for charm meson. The first excited states $0^{+}$ and
$1^{+}$ heavy mesons are the quantum numbers of the
$s_{l}^{p}=\frac{1}{2}^{+}$ doublet. There is also an excited
doublet of heavy meson with $J^{P}=1^{+}$ and $2^{+}$. Similarly,
other excited mesonic states have their $J^{P}$ states. In this
article, we identify the recent charmed meson states $D_{J}(2550)$,
 $D_{J}^{\ast}(2600)$, $D_{J}(2740)$,
$D_{J}^{\ast}(2760)$, $D_{J}(3000)$ and $D_{J}^{\ast}(3000)$ with
their $J^{P}$ assignment. These states were observed by LHCb
collaboration and predicting their decay widths and masses. We study
strong decays of these charmed mesons to ground states heavy mesons
along with the emission of pseudo-scalar pions in heavy quark
effective theory in the leading order approximations. Although the
same work has been studied by \cite{Wang} but we extend their
predictions by fitting the experimental data to find the coupling
constants in various strong decays. Also, we include two additional
possibilities for assignment of $J^{P}$ states to
$D_{J}^{\ast}(3000)$ and $D_{J}(3000)$. In the end, we also try to
justify all the possible assignments to $D_{J}^{\ast}(3000)$ and
$D_{J}(3000)$ by analyzing their branching ratios graphically.
\section{The Lagrangian for Strong Decays to Heavy Mesons}
A single field $H_{a}$ where it annihilates the
$s_{l}=\frac{1}{2}^{-}$ meson doublet, pseudoscalar and vector
mesons can be mentioned as\cite{Falk92}
\begin{equation}
H_{a}=\frac{1+\slashed{v}}{2}(P_{a}^{\ast\mu}\gamma_{\mu}-P_{a}\gamma_{5})
\end{equation}
Here a is the SU(3) index. In charm mesons sector, $H_{a}$ consists
of the $D^{0},D^{+},D_s^{+}$ pseudo-scalar mesons and $D^{*0},
D^{*+}, D_s^{*+}$ vector mesons. The lowest lying excited states are
the $J^{P}=0^{+}$  and $1^{+}$ i.e. $s_{l}^{P}=\frac{1}{2}^{+}$
doublet and represented by the fields $S_{a}$ \cite{Falk}. The
fields for excited spin doublets are mentioned below:
\begin{equation}
S_{a}=\frac{1+\slashed{v}}{2}(P_{1a}^{\mu}\gamma_{\mu}\gamma_{5}-P_{0a}^{\ast})
\end{equation}
\begin{equation}
T_{a}^{\mu}=\frac{1+\slashed{v}}{2}(P^{\ast\mu\nu}_{2a}\gamma_{\mu}-P_{1a\nu}\sqrt{\frac{3}{2}}\gamma_{5}[g^{\mu\nu}-\frac{1}{3}\gamma^{\nu}(\gamma^{\mu}-v^{\mu})])
\end{equation}
\begin{equation}
X_{a}^{\mu}=\frac{1+\slashed{v}}{2}(P^{\ast\mu\nu}_{2a}\gamma_{5}\gamma_{\nu}-P^{\prime\ast}_{1a\mu}\sqrt{\frac{3}{2}}[g^{\mu\nu}-\frac{1}{3}\gamma^\nu(\gamma^{\mu}+v_{\mu})])
\end{equation}
\begin{equation}
Y_{a}^{\mu\nu}=\frac{1+\slashed{v}}{2}(P_{3a}^{\ast\mu\nu\sigma}\gamma_{5}\gamma_{\sigma}-P_{2a}^{\alpha\beta}\frac{5}{3}\gamma_{5}[g_{\alpha}^{\mu}g_{\beta}^{\nu}-\frac{g_{\beta}^{\nu}\gamma_{\alpha}(\gamma^{\mu}-v^{\mu})}{5}-\frac{g_{\alpha}^{\mu}\gamma_{\beta}(\gamma^{\nu}-v^{\nu})}{5}]
\end{equation}
\begin{equation}
Z_{a}^{\mu\nu}=\frac{1+\slashed{v}}{2}(P_{3a}^{\ast\mu\nu\sigma}\gamma_{5}\gamma_{\sigma}-P_{2a}^{\ast\alpha\beta}\sqrt{\frac{5}{3}}[g_{\alpha}^{\mu}g_{\beta}^{\nu}-\frac{g_{\beta}^{\nu}\gamma_{\alpha}(\gamma^{\mu}+v^{\mu})}{5}-\frac{g_{\alpha}^{\mu}\gamma_{\beta}(\gamma^{\nu}+v^{\nu})}{5}])
\end{equation}
\begin{equation}
R_{a}^{\mu\nu\rho}=\frac{1+\slashed{v}}{2}(P_{4a}^{\ast\mu\nu\rho\sigma}\gamma_{5}\gamma_{\sigma}-P_{3a}^{\alpha\beta\tau}\sqrt{\frac{7}{4}}[g_{\alpha}^{\mu}g_{\beta}^{\nu}g_{\tau}^{\rho}-\frac{g_{\beta}^{\nu}g_{\tau}^{\rho}\gamma_{\alpha}(\gamma^{\mu}-v^{\mu})}{7}-\frac{g_{\alpha}^{\nu}g_{\tau}^{\rho}\gamma_{\beta}(\gamma^{\mu}-v^{\mu})}{7}-\frac{g_{\alpha}^{\mu}g_{\beta}^{\nu}\gamma_{\tau}(\gamma^{\rho}-v^{\rho})}{7}])
\end{equation}
The super fields $H_{a}$ contain s-wave mesons whereas $S_{a},T_{a}$
contain the p-wave mesons. The light pseudoscalar mesons are
described by the fields $\xi=\exp^{\frac{iM}{f}}$. The pion octet is
introduced by the vector and axial combinations
$V^{\mu}=\frac{1}{2}{\xi\partial^{\mu}\xi^{\dag}+\xi^{\dag}\partial^{\mu}\xi}$
and
$A^{\mu}=\frac{1}{2}{\xi\partial^{\mu}\xi^{\dag}-\xi^{\dag}\partial^{\mu}\xi}$
We choose $f_{\pi}=130MeV$. Here, all traces are taken over Dirac
spinor indices, light quark $SU(3)_{V}$ flavor indices $a=u, d, s$
and heavy quark flavor indices $Q=c,b$. The Dirac structure of
chiral Lagrangian has been replaced by velocity vector $v_{µ}$. At
the leading order, the heavy meson chiral chiral Lagrangian terms
$L_{H}$,$L_{S}$,$L_{T}$,$L_{X}$,$L_{Y}$,$L_{Z}$,$L_{R}$ for the
strong decays to the $D^{(\ast)}\pi$,$D^{(\ast)}\eta$ and
$D_{s}^{(\ast)}K$ states can be written as:
\begin{equation}
L_{H}=g_{H}Tr\{\bar{H_{a}}H_{b}\gamma_{\mu}\gamma_{5}\emph{A}^{\mu}_{ba}\}
\end{equation}
\begin{equation}
L_{S}=g_{S}Tr\{\bar{H_{a}}S_{b}\gamma_{\mu}\gamma_{5}\emph{A}^{\mu}_{ba}\}+h.c.,
\end{equation}
\begin{equation}
L_{T}=\frac{g_{T}}{\Lambda}Tr(\bar{H_{a}}T_{b}^{\mu}(\emph{D}_{\mu}\slashed{\emph{A}}+i\slashed{D}\emph{A}_{\mu})_{ba}\gamma_{5})+h.c.,
\end{equation}
\begin{equation}
L_{X}=\frac{g_{X}}{\Lambda}Tr(H_{a}X_{b}^{\mu}(i\emph{D}_{\mu}\slashed{\emph{A}}+i\slashed{\emph{D}}\emph{A}_{\mu})_{ba}\gamma_{5})+h.c.,
\end{equation}
\begin{equation}
L_{Y}=\frac{1}{\Lambda^{2}}Tr(\bar{H_{a}}Y_{b}^{\mu\nu}[k_{1}^{Y}\{\emph{D}_{\mu},\emph{D}_{\nu}\}\emph{A}_{\lambda}+k_{2}^{Y}(\emph{D}_{\mu}\emph{D}_{\lambda}\emph{A}_{\mu}+\emph{D}_{\nu}\emph{D}_{\lambda}\emph{A}_{\mu})]_{ba}\gamma^{\lambda}\gamma_{5})+h.c.,
\end{equation}
\begin{equation}
L_{Z}=\frac{1}{\Lambda^{2}}Tr(\bar{{H_{a}}}Z_{b}^{\mu\nu}[k_{1}^{z}\{\emph{D}_{\mu},\emph{D}_{\nu}\},\emph{D}_{\nu}\}\emph{A}_{\lambda}+k_{2}^{Z}(\emph{D}_{\mu},\emph{D}_{\lambda}\emph{A}_{\nu}+\emph{D}_{\nu}\emph{D}_{\lambda}\emph{A}_{\mu})]_{ba}\gamma^{\lambda}\gamma_{5})+h.c.,
\end{equation}
where $\emph{D}_{\mu}=\partial_{\mu}+\nu_{\mu}$,
$\{\emph{D}_{\mu},\emph{D}_{\nu}\}=\emph{D}_{\mu}\emph{D}_{\nu}+\emph{D}_{\mu}\emph{D}_{\nu}$,$\{\emph{D}_{\mu},\emph{D}_{\nu},\emph{D}_{\rho}\}=\emph{D}_{\mu}\emph{D}_{\nu}\emph{D}_{\rho}+\emph{D}_{\mu}\emph{D}_{\rho}\emph{D}_{\nu}+\emph{D}_{\nu}\emph{D}_{\rho}\emph{D}_{\mu}+\emph{D}_{\rho}\emph{D}_{\mu}\emph{D}_{\nu}+\emph{D}_{\rho}\emph{D}_{\nu}\emph{D}_{\mu}$,
These terms describe the transitions of positive and negative parity
mesons with the emission of light pseudo-scalar mesons. The mixing
angle between two states are determined by including spin symmetry
violating corrections in the Lagrangian. The term should respect
parity and time reversal and may be of generic form as written
below.
\begin{equation}
\emph{L}_{d1}=\frac{h_{1}}{2m\Lambda}Tr[\bar{H}\sigma^{\mu\nu}T^{\alpha}\sigma_{\mu\nu}\gamma^{k}\gamma^{5}(iD_{\alpha}A_{\kappa}+iD_{\kappa}A_{\alpha})]+h.c.
\end{equation}
The corresponding operator for the mixing of $1^{+}$ in 2S and 1D
respectively is due to spin symmetry violating effect and can be
written as:
$\emph{L}_{mix}=g_{1}Tr[\bar{\tilde{H}}\phi^{\mu\nu}_{s}X_{\mu}\sigma_{\nu\alpha}\emph{v}^{\alpha}]+h.c.$.

\section{Strong Decay Width Formula and Coupling Constants}
From the chiral Lagrangian terms, we can obtain the decay widths
$\Gamma$ for strong decays to final states $D^{(\ast)}\pi$,
$D^{(\ast)}\eta$, $D_{s}^{(\ast)}K$ where the symmetry breaking
scale $\Lambda_{X}=1GeV$. The expression for decay widths if we
consider various doublets which the decaying meson belongs to are as
follows where M denotes the emission of pseudoscalar mesons i.e.
$\pi$, K and $\eta$ fields.
\\ $(0^{-},1^{-})$ to $(0^{-},1^{-})+M$\\
\begin{equation}
\Gamma(1^{-}\rightarrow0^{-})=C_{M}\frac{g_{H}^{2}}{6\pi
f_{\pi}^{2}}\frac{M_{f}}{M_{i}}\mid\vec{p_{M}}\mid^{3}
\end{equation}
\begin{equation}
\Gamma(1^{-}\rightarrow1^{-})=C_{M}\frac{g_{H}^{2}}{3\pi
f_{\pi}^{2}}\frac{M_{f}}{M_{i}}\mid\vec{p_{M}\mid}^{3}
\end{equation}
\\
$(0^{+},1^{+})$ to $(0^{-},1^{-})+M$\\
\begin{equation}
\Gamma(1^{+}\rightarrow 1^{-})=C_{M}\frac{g_{S}^{2}}{2\pi
f_{\pi}^{2}}\frac{M_{f}}{M_{i}}\mid\vec{p_{M}\mid}[m_{M}^{2}+\mid\vec{p_{M}}\mid^{2}]
\end{equation}
\begin{equation}
\Gamma(1^{+}\rightarrow 0^{-})=C_{M}\frac{g_{S}^{2}}{2\pi
f_{\pi}^{2}}\frac{M_{f}}{M_{i}}\mid
\vec{p_{M}\mid}[m_{M}^{2}+\mid\vec{p_{M}}\mid^{2}
\end{equation}
\\
$(1^{+},2^{+})$ to $(0^{-},1^{-})+M$
\begin{equation}
\Gamma(1^{+}\rightarrow 1^{-})=C_{M}\frac{2g_{T}^{2}}{3\pi
f_{\pi}^{2}\Lambda^{2}}\frac{M_{f}}{M_{i}}\mid\vec{p_{M}}^{5}\mid
\end{equation}
\begin{equation}
\Gamma(2^{+}\rightarrow 0^{-})=C_{M}\frac{4g_{T}^{2}}{15\pi
f_{\pi}^{2}\Lambda^{2}}\frac{M_{f}}{M_{i}}\mid\vec{p_{M}}^{5}\mid
\end{equation}
\begin{equation}
\Gamma(2^{+}\rightarrow 1^{-})=C_{M}\frac{2g_{T}^{2}}{5\pi
f_{\pi}^{2}\Lambda^{2}}\frac{M_{f}}{M_{i}}\mid\vec{p_{M}}^{5}\mid
\end{equation}
\\
$(1^{-},2^{-})$ to $(0^{-},1^{-})+M$
\begin{equation}
\Gamma(1^{-}\rightarrow 0^{-})=C_{M}\frac{4 g_{X}^{2}}{9\pi
f_{\pi}^{2}\Lambda^{2}}\frac{M_{f}}{M_{i}}\mid
\vec{p_{M}}^{3}\mid[m_{M}^{2}+\mid\vec{p_{M}}\mid^{2}]
\end{equation}
\begin{equation}
\Gamma(1^{-}\rightarrow 1^{-})=C_{M}\frac{2 g_{X}^{2}}{9\pi
f_{\pi}^{2}\Lambda^{2}}\frac{M_{f}}{M_{i}}\mid\vec{p_{M}}^{3}\mid[m_{M}^{2}+\mid\vec{p_{M}}\mid^{2}]
\end{equation}
\begin{equation}
\Gamma(2^{-}\rightarrow 1^{-})=C_{M}\frac{2 g_{X}^{2}}{3\pi
f_{\pi}^{2}\Lambda^{2}}\frac{M_{f}}{M_{i}}\mid\vec{p_{M}}^{5}\mid[m_{M}^{2}+\mid\vec{p_{M}}\mid^{2}]
\end{equation}
\\
$(2^{-},3^{-})$ to $(0^{-},1^{-})+M$
\begin{equation}
\Gamma(2^{-}\rightarrow 1^{-})=C_{M}\frac{4g_{Y}^{2}}{15\pi
f_{\pi}^{2}\Lambda^{4}}\frac{M_{f}}{M_{i}}\mid\vec{p_{M}}^{7}\mid
\end{equation}
\begin{equation}
\Gamma(3^{-}\rightarrow 0^{-})=C_{M}\frac{4g_{Y}^{2}}{35\pi
f_{\pi}^{2}\Lambda^{4}}\frac{M_{f}}{M_{i}}\mid\vec{p_{M}}^{7}\mid
\end{equation}
\begin{equation}
\Gamma(3^{-}\rightarrow 1^{-})=C_{M}\frac{16g_{Y}^{2}}{105\pi
f_{\pi}^{2}\Lambda^{4}}\frac{M_{f}}{M_{i}}\mid\vec{p_{M}}^{7}\mid
\end{equation}
The coefficients $C_{M}$ are different for the various light
pseudoscalar mesons: $C_{\pi^{+}}=C_{K^{+}}=1,C_{\pi^{0}}=
C_{K_{s}}=\frac{1}{2},C_{\eta}=\frac{1}{6}$. $\vec{p_{M}}$ is the
three momentum of M. The higher order corrections to heavy quark
limit can also be considered by adding terms of the order
$\frac{1}{m_{Q}}$ with some un-known constants. The decay rates
depend upon effective coupling constants. The parameters used in the
above expressions for decay widths are taken from the particle data
group\cite{Beringer}. Thus the numerical values of decay widths
comes out in terms of coupling constants $g_{H},g_{Y},g_{X}$ etc.
Here the first radial excitation of $D^{\ast}$ is represented as
$\tilde{D}^{\ast}$. The first radially excited state of H is
governed by the decay constant $\tilde{g_{H}}$ which can be fitted
to experimental data within mass range of 2600-2700 MeV. Coupling
constants can either be determined theoretically or by fitting the
experimental data. However, various quark models
\cite{Donnell,Becirevic} and sum rule (eg. QCD sum
rules)\cite{Colangelo1,Colangelo3,Grozin} techniques predict the
coupling constants. Another possible method is to use lattice QCD
\cite{Divittis} which incorporate QCD in their first principle.
Using experimental data of decay widths and branching ratios as
input, one can fit the experimental data to find the effective
coupling constants. The coupling constants play an important role in
heavy quark phenomenology. They are directly related to charm meson
strong decays and are further useful to explore other decays of
charm mesons involving pionic emissions.
\section{Spin-parity Analysis for Non-strange Charm Meson States}
The recent experimental data of charm meson states from LHCb and
BaBar collaboration motivates us to find the best fit values of
coupling constants in strong decays.
The table 1 mentions the recent experimental data of non-strange charm mesons.\\
\begin{center}
\label{Table 1}\captionof{table}{~Experimental Status of Latest
Charm Mesons} \noindent\hspace*{-1.0cm}\scalebox{0.7}{
\begin{tabular}{l l l l l l l}
\hline
Sr.No & Charm Meson state&Mass[MeV][LHCb]&Mass[MeV][BaBar]&Width[MeV][LHCb]&Width[MeV][BaBar]&Decay Channel\\
1&$D_{J}^{\ast}(2650)^{0}$&$2649.2\pm3.5\pm3.5$&$2608.7\pm2.4\pm2.5$&$140.2\pm17.1\pm18.6$&$93\pm6\pm13$&$D^{\ast+}\pi^{-}$\\
2&$D_{J}^{\ast}(2760)^{0}$&$2761.1\pm5.1\pm6.5$&$2763.3\pm2.3\pm2.3$&$74.4\pm3.4\pm37.0$&$60.9\pm5.1\pm3.6$&$D^{\ast+}\pi^{-}$\\
3&$D_{J}(2580)^{0}$&$2579.5\pm3.4\pm5.5$&$2539.4\pm4.5\pm6.8$&$177.5\pm17.8\pm46.0$&$130\pm12\pm13$&$D^{\ast+}\pi^{-}$\\
4&$D_{J}(2740)^{0}$&$2737.0\pm3.5\pm11.2$&$2752.4\pm1.7\pm2.7$&$73.2\pm13.4\pm25.0$&$71\pm6\pm11$&$D^{\ast+}\pi^{-}$\\
5&$D_{J}(3000)^{0}$&$2971.8\pm8.7$&$188.1\pm44.8$&&&$D^{\ast+}\pi^{-}$\\
\hline\\
6&$D_{J}^{\ast}(2760)^{0}$&$2760.1\pm1.1\pm3.7$&$74.4\pm3.4\pm19.1$&&&$D^{+}\pi^{-}$\\
7&$D_{J}^{\ast}(3000)^{0}$&$3008.1\pm4.0$&$110.5\pm11.5$&&&$D^{+}\pi^{-}$\\
\hline\\
8&$D_{J}^{\ast}(2760)^{+}$&$2771.7\pm1.7\pm3.8$&$66.7\pm6.6\pm10.5$&&&$D^{0}\pi^{+}$\\
9&$D_{J}^{\ast}(3000)^{+}$&3008.1&$66.7\pm6.6\pm10.5$&&&$D^{0}\pi^{+}$\\
\hline
\end{tabular}
}
\end{center}
 The states with
$J^{P}=(0^{-},1^{-},0^{+},1^{+},1^{+},2^{+})$ are well known. The
doublets having spin-parity assignments $s_{l}^{P}=\frac{3}{2}^{+}$,
consists of $D_{1}(2420)$ and $D_{2}^{\ast}(2460)$ in non-strange
sector. The states $(D_{0}^{\ast}(2400),D_{1}^{\prime}(2430)$ belong
to $s_{l}^{P}=\frac{1}{2}^{+}$ charm doublet\cite{Sanchez}. The
experimental data on decay widths suggest that states
$(0^{+},1^{+})$ are quite broad, expecting to decay via s-wave
whereas the states belonging to $(1^{+},2^{+})$ doublets are quite
narrow and decay via d-wave. The measured branching ratio by BaBar
Collaboration is given as:
\begin{equation}
\frac{BR(D_{2}^{\ast0}(2460)\rightarrow
D^{+}\pi^{-})}{BR(D_{2}^{\ast0}(2460)\rightarrow
D^{\ast+}\pi^{-})}=1.47\pm0.03\pm0.16
\end{equation}
There are few more recent states whose branching ratios as measured
by BaBar is mentioned below.
\begin{equation}
\frac{BR(D^{0}(2600)\rightarrow
D^{+}\pi^{-})}{BR(D^{0}(2600)\rightarrow D^{\ast+}
\pi^{-})}=0.32\pm0.02\pm0.09
\end{equation}

\begin{equation}
\frac{BR(D^{0}(2760)\rightarrow
D^{+}\pi^{-})}{BR(D^{0}(2750)\rightarrow
D^{\ast+}\pi^{-})}=0.42\pm0.05\pm0.11
\end{equation}

The information from the BaBar Collaboration and the quark model
suggests that $D^{0}(2550)$ state lies in $0^{-}$ state. The
$D^{0}(2600)$ corresponds to $1^{-}$ state either in the 2S or 1D
spectrum respectively because this state was observed in both
$D{\pi}$ and $D^{\ast}\pi$ channels. If we find the mass of these
particular states using heavy quark symmetry and other theoretical
models \cite{Ebert},\cite{Pierro}, it can be suggested that the
state $D(2600)$ can be identified as either radial excitation of
heavy quark doublet $H$ or 1D. The branching ratios for $D\pi$ and
$D^{\ast}\pi$ for both the decay states are calculated as
$\frac{BR(D^{0}(2600)\rightarrow
D^{+}\pi^{-})}{BR(D^{0}(2600)\rightarrow D^{\ast+}\pi^{-})}= 0.82$
for 2S and $\frac{BR(D^{0}(2600)\rightarrow
D^{+}\pi^{-})}{BR(D^{0}(2600)\rightarrow D^{\ast+}\pi^{-})}=0.38$
The comparison with the experimental data results in favor of 1D
assignment. Therefore, the possible assignment for this particular
state can be 1D respectively. The theoretical estimation of coupling
constant for the strong decay width of mesons in this particular
state is $0.53\pm0.01$. The theoretical estimation of branching
ratios from the heavy quark effective theory \cite{Colangelo} leads
to the conclusion that there may be possibility of violations in
flavor and spin symmetry. In ref.\cite{Sun}, Sun et al. used the
$^{3}P_{0}$ model to examine the strong decays of these states and
they concluded that $D^{0}(2600)$ state can be identified as the
mixture of $2^{3}S_{1}$ and $1^{3}D_{3}$ state. Therefore, the other
possibility is that D(2600) may be considered as a mixing state of
2S and 1D respectively.
The other two states $D(2750)$ and $D(2760)$ can be identified with
$J^{P}=(2^{-},3^{-})$ assignment. It is very interesting to point
out that non-strange partner of $D_{sJ}(2860)$ can be associated
with D(2760) due to mass gap which is about 150 MeV. There are also
several references like \cite{Wang},\cite{Colangelo} which suggest
possible assignment for $D(2750)$ and $D^{\ast}(2760)$ state with
the l=2, n=1 state. Moreover, the branching ratio measurement
$\frac{BR(D^{0}(2760)\rightarrow
D^{+}\pi^{-})}{BR(D^{0}(2750)\rightarrow D^{\ast+}\pi^{-})}$ gives
value 0.80 from the leading order effective theory which is found to
be matching with the experimental data. Saturating the total decay
width with the ground state to two body decays, we can fit the
experimental data of LHCb and BaBar to estimate the coupling
constant. We take the experimental data of decay widths of recent
states as mentioned in table 1 to find the hadronic coupling
constants. The decay widths are calculated using the decay formulae
given above. We can fit the experimental data of BaBar and LHCb
collaborations to estimate the coupling constants. The observed
radially excited non-strange charm meson states in the heavy meson
spectrum are the two resonances (D(2550),$D^{\ast}(2600)$). The
values of coupling constant is obtained from the measured width of
(D(2550)) and the computed value is $0.35\pm0.03$ \cite{Colangelo}.
The calculated value of coupling constants in our fitting program
for D(2550) comes out to be $0.40\pm0.05$. The best fit value of
these coupling constants is estimated with in experimental error
using chi-square minimization technique. The errors are clearly
dominated by the statistical and systematic uncertainties. To check
the consistency of our fitting algorithm, the coupling constant
estimation is being carried out for the decays of
$J^{P}=(0^{+},1^{+})$ and the fitted value comes out to be
$0.53\pm0.04$. This value has been found to be matching well with
predictions from other theoretical approaches\cite{Colangelo}. The
value of coupling constants for D(2750) and $D^{\ast}(2760)$ states
are also estimated in our fitting program which are $0.61\pm0.01$
and $0.79\pm0.03$. Let us consider new states predicted by LHCb in
the $D\pi$ and $D^{\ast}\pi$ spectrum and from the strong decays,the
states are labeled as $D_{J}^{\ast}(3000)^{0}\rightarrow
D^{+}\pi^{-}$ and $D_{J}^{\ast}(3000)^{+}\rightarrow D^{0}\pi^{+}$.
On the basis of LHCb data\cite{Sanchez},the angular distribution of
$D_{J}(3000)\rightarrow D^{\ast}\pi$ is found to be consistent with
unnatural parity. The possible spin-parity assignment for
$D_{J}(3000)^{0}$ can be $J^{P}=0^{-},1^{+},2^{-},3^{+}...$ and
$D_{J}^{\ast}(3000)$ have possible spin-parity assignments
$J^{P}=0^{+},1^{-},2^{+},3^{-},4^{+}.....$. Thus, it can be stated
that the two states can be higher radial excitations or can belong
to 1F states in the meson spectrum. Some possible indications about
the possible assignment of $J^{P}$ quantum numbers can be realized
from the masses of these states. One of the well known potential
models \cite{Ebert} calculated the masses of all possible excited
mesonic states and we can suggest the following possible spin and
parities assignments of these newly discovered states. To extract
the detailed information about the newly observed states, we present
the summary of all D meson states and various possibilities of
$D_{J}^{\ast}(3000)$ and $D_{J}^{\ast}(3000)^{0}$ in the table 2
below.
\begin{center}
\label{Table 2} \captionof{table}{~Table showing all non-strange
charm meson states}
\noindent\hspace*{-2.0cm}\scalebox{0.6}{
\begin{tabular}{l l l l l l l l}
\hline\\
$s_{l}^{p}$&$\frac{1}{2}^{-}$&$\frac{1}{2}^{+}$&$\frac{3}{2}^{+}$&$\frac{3}{2}^{-}$&$\frac{5}{2}^{+}$&$\frac{7}{2}^{+}$&$\frac{5}{2}^{-}$\\
\hline\\
n=1\\
&$D(1869)(J^{P}=0^{-})$&$D_{0}^{\ast}(2400)(J^{P}=0^{+})$&$D_{1}(2460)(J^{P}=1^{+})$&$D(2750)\cite{Colangelo}(J^{P}=1^{-})$&$D_{J}^{\ast}(3000)(J^{P}=2^{+})\star$&$D_{J}(3000)(J_{P}=3^{+})\star$&$D_{J}(3000)(J^{P}=2^{-})\star$\\
&$D^{\ast}(2010)(J^{P}=1^{-})$&$D_{1}^{\backprime}(2430)(J^{P}=1^{+})$&$D_{2}^{\ast}(2430)(J^{P}=2^{+})$&$D(2760)\cite{Colangelo}(J^{P}=2^{-})$&$D_{J}(3000)(J^{P}=3^{+})\star$&$D_{J}^{\ast}(3000)(J_{P}=4^{+})\star$&$D_{J}^{\ast}(3000)(J^{P}=3^{-})\star$\\
\hline\\
n=2\\
&$D(2550)(J^{P}=0^{-})$&$D_{J}^{\ast}(3000)(J^{P}=0^{+})\star$&$D_{J}(3000)(J^{P}=1^{+})\star$&$D_{J}^{\ast}(3000)(J^{P}=1^{-})\star$&&&\\
&$D(2600)(J^{P}=1^{-})$&$D_{J}(3000)(J^{P}=1^{+})\star$&$D_{J}^{\ast}(3000)(J^{P}=2^{+})\star$&$D_{J}(3000)(J^{P}=2^{-})\star$&&&\\
\hline\\
n=3\\
&$D_{J}(3000)(J^{P}=0^{-})\star$&$D_{J}^{\ast}(3000)(J^{P}=1^{-})\star$&&&&&\\
\hline\\
\end{tabular}
}
\end{center}
Wang et al.\cite{Wang} also suggested the various possibilities of
$D_{J}(3000)$ states and calculated the decay widths of these states
in terms of relevant coupling constants. We also suggest the same
but we add the two more possibilities i.e. $(1^{-},2^{-})$ and
$(2^{-},3^{-})$ lying in 1D spectrum. To analyze the spectrum of
above $J^{P}$ states, we study the two body decay behavior and
calculate the branching ratios for the states for which decay to $P
M$ and $P^{\ast}M$ both are allowed. 
\begin{center}
\captionof{table}{~Ratios of decay width for $D_{J}^{\ast}(3000)$
state} \label{Table 3}
\hspace*{-0.5cm}
\begin{tabular}{l l l}\hline
$D_{J}^{\ast}(3000)$&$D_{J}^{\ast}(3000)\rightarrow
D^{\ast}\pi$&$\frac{BR(D_{J}^{\ast}\rightarrow
D^{\ast+}\pi^{-})}{BR(D_{J}^{\ast}\rightarrow D^{+}\pi^{-})}$\\
\hline\\
$s_{l}^{p}=\frac{5}{2}^{+},J^{P}=2^{+}$&f-wave&0.343\\
$s_{l}^{p}=\frac{7}{2}^{+},J^{P}=4^{+}$&f-wave&0.52\\
$s_{l}^{p}=\frac{1}{2}^{+},J^{P}=0^{+}$&s-wave&0\\
$s_{l}^{p}=\frac{3}{2}^{+},J^{P}=2^{+}$&d-wave&0.955\\
$s_{l}^{p}=\frac{1}{2}^{-},J^{P}=1^{-}$&p-wave&1.57\\
$s_{l}^{p}=\frac{5}{2}^{-},J^{P}=3^{-}$&f-wave&0.68\\
$s_{l}^{p}=\frac{3}{2}^{-},J^{P}=1^{-}$&p-wave&0.32\\
\hline
\end{tabular}
\end{center}
The table 3 collects the ratios of decay width for
$D_{J}^{\ast}(3000)$ state. Predicted ratios along with graphs can
be analyzed to exclude some of the assignments.
\begin{figure*}
\vspace{-0.5 cm}
\includegraphics[width=0.55\columnwidth,clip=true]{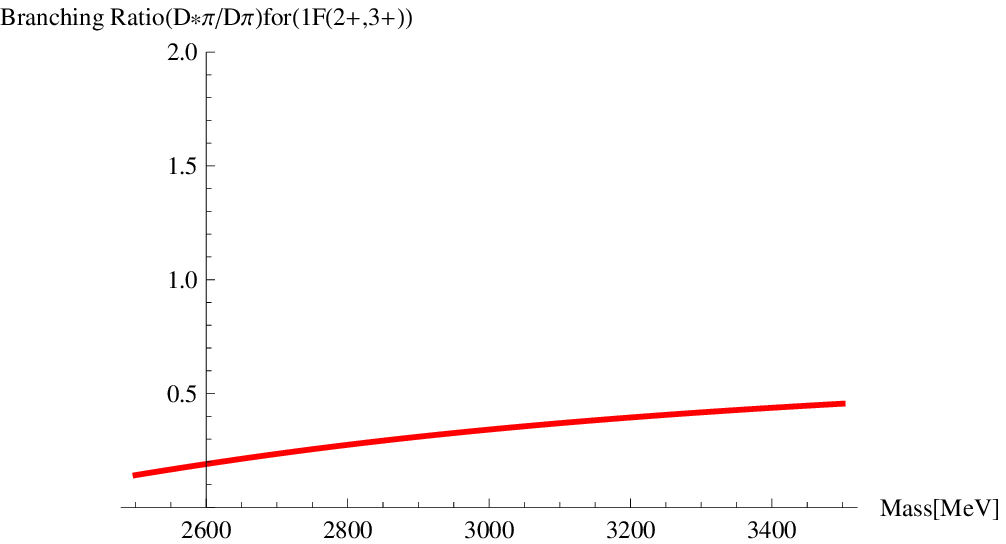}
\hspace{-0.3 cm}
\includegraphics[width=0.55\columnwidth,clip=true]{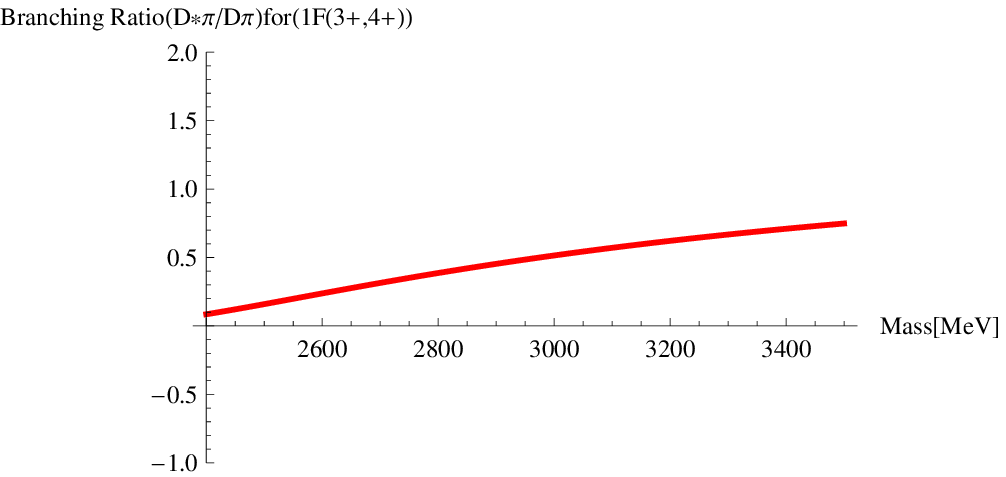}
\hspace{-0.3 cm}
\includegraphics[width=0.55\columnwidth,clip=true]{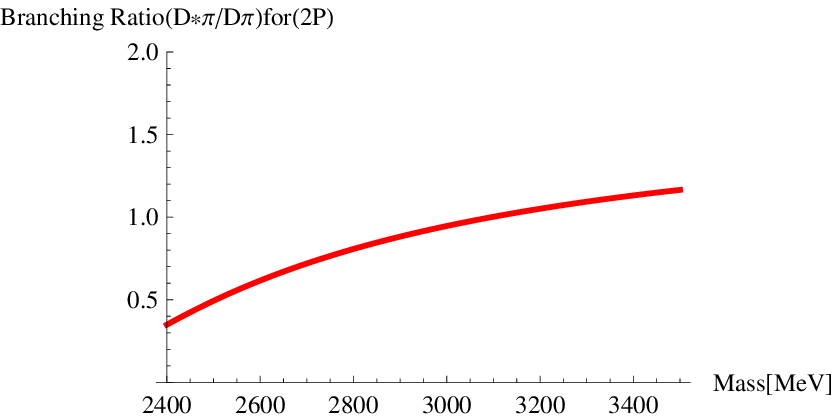}
\hspace{-0.3 cm}
\includegraphics[width=0.55\columnwidth,clip=true]{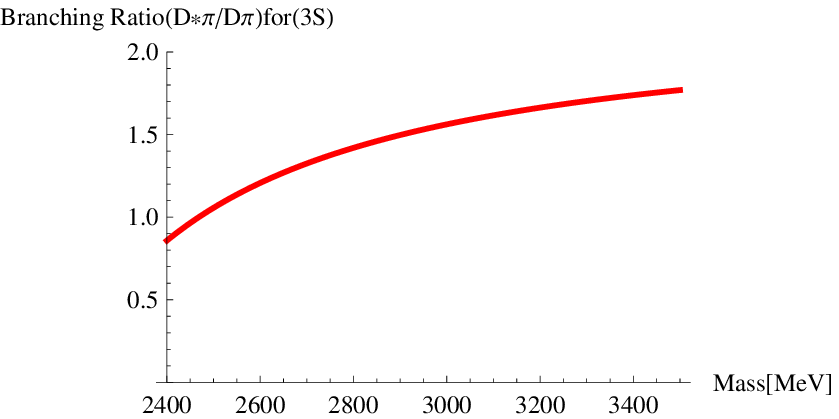}
\hspace{-0.3 cm}
\includegraphics[width=0.55\columnwidth,clip=true]{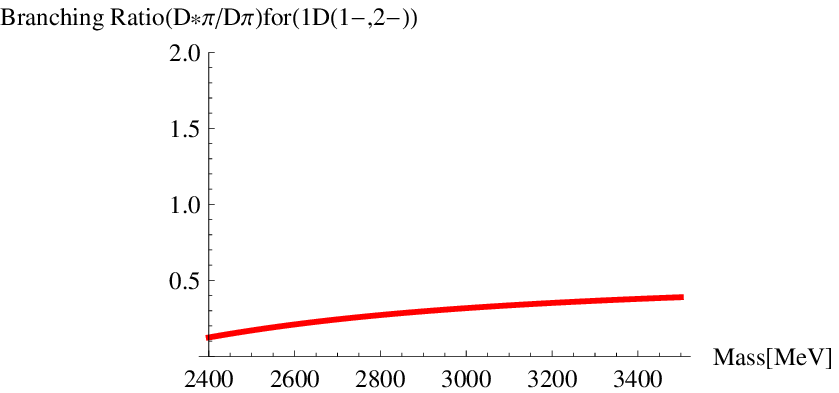}
\hspace{-0.3 cm}
\includegraphics[width=0.55\columnwidth,clip=true]{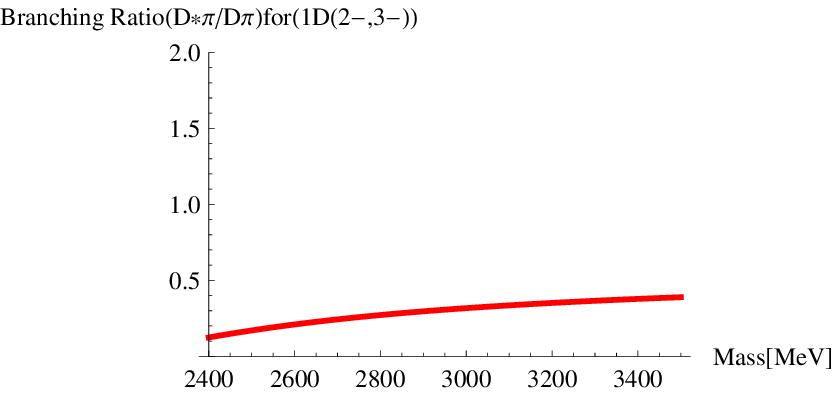}
\hspace{-0.3cm}\label{Fig. 1} \caption{~Graph showing branching
ratios Vs mass of decaying particle}
\end{figure*}
\\
The graphs show the variation of branching ratios Vs mass of
decaying particle. From the graphs, it can be stated that the states
lying in 3S doublet around the mass values of 3000 MeV does not
produce ratios less than or around 1.
The spin partner of $D_{J}^{\ast}(3000)$ is $D_{J}(3000)$ which is
observed to decay via $D^{+}\pi^{-}$ channel. The other possible
decay modes are $DK$ and $D\eta$. If $D_{J}(3000)$ and
$D_{J}^{\ast}(3000)$ belongs to $2P(0^{+},1^{+})$ doublet then the
allowed decay modes for $D_{J}^{\ast}(3000)$ does not include
$D^{\ast+}\pi^{-}$ decay channels which agrees well with the
experiments. In all other cases, decays to both the channels $D\pi$
and $D^{\ast}\pi$ are observed. Moreover, the calculation of
branching ratios ($\frac{D_{J}^{\ast}(3000)\rightarrow
D^{+}\pi^{-}}{D_{J}^{\ast}(3000)\rightarrow D^{\ast+}\pi^{-}}$) for
all possible doublets suggests that the possible assignment of these
two states can also be either ($J^{P}=(1^{-},2^{-})$) or
($J^{P}=(2^{+},3^{+})$) for $D_{J}^{\ast}(3000)$ and $D_{J}(3000)$
mesons. 
The state $3^{3}S_{1}$ state decays to
$D^{\ast}\pi$,$D\pi$,$D_{s}^{\ast}K$,$D_{s}K$ and $D\eta$ and the
coupling constant for this particular state lies near the value of
$\simeq0.1$. Also, the similar state appears to be a narrow D meson
state. But its partial decay width to $D^{\ast}\pi$ appears to be
almost double than that of $D\pi$ mode which is completely
inconsistent with predictions by LHCb collaboration. If we assign
$2P(J^{P}=(1^{+},2^{+}))$ doublet to these mesons than the coupling
constant for strong decays to $\pi,K,\eta$ should lie between
0.12-0.15. For ($2^3D_{3}$) state, $(D\pi,DK,D\eta)$ are the allowed
decay modes. Additional decay channels may include $D(2460)\pi$ and
$D(2420)\pi$. The additional information about these decay channels
may help to estimate coupling constant precisely. In addition to
this, $2^3D_{1}$ state has most prominent decay mode is $D\pi$
therefore the ideal decay mode to search for this state is $D\pi$.
The ratio of total decay widths for $D_{J}^{\ast}(3000)$ and
$D_{J}(3000)$ can be calculated from the experimental data of LHCb
collaboration.
\begin{equation}
\frac{\Gamma(D_{J}^{\ast}(3000))}{\Gamma(D_{J}(3000))}=0.587\pm0.083
\end{equation}
The strong decay width ratios when calculated in HQET with leading
order for all the possible assignments produce values very far from
the experimental values. Thus, only $\frac{1}{m_{Q}}$ corrections
when included, more authentic results can be produced. Similar
analysis can be carried out for the $D_{J}(3000)$ mesons. The
various possibilities include $3^{1}S_{0}$, $2P(1^{+})$ and
$1D(2^{-})$ and 1F etc.. A closer look at the decay width of all the
above states in the $D^{\ast}\pi$ spectrum suggests that the most
possible assignment can belong to 2P state. The most prominent decay
mode for $2P(1^{+})$ in $(1^{+},2^{+})$ doublet is found to be
$D^{\ast}\pi$ which matches well with the experimental data on decay
width for $D_{J}(3000)$. However, if we consider the $D_{J}(3000)$
as the spin partner of $D_{J}^{\ast}(3000)$ then the only
possibility can be that 2P$(1^{+})$ in $(0^{+},1^{+})$ overlaps with
that with $1^{+}$ in $(1^{+},2^{+})$ doublet.
\section{Conclusion}
We study the heavy meson decay width in the framework of heavy quark
effective theory that represents heavy quark and chiral symmetry at
$\Lambda_{QCD}\simeq 1GeV$. We studied the recent charm meson states
with their $J^{p}$ assignment. The upcoming results at
collaborations like LHCb produces the data for branching ratios that
is used to calculate the decay width, coupling constants and
suitable $J^{p}$ states. The coupling constants and their studies
are important to study heavy meson phenomenology. The accurate
prediction of coupling constants help to study the detailed
interaction of heavy mesons. The present work calculates the
coupling constant for strong decays of non-strange charm meson
states (D(2550),D(2600),D(2750)and D(2760)) by using Chi-square
minimization techniques. The numerical values of decay widths from
the collaborations like LHCb, BaBaR and CDF are used to extract the
values of coupling constants. The various assignments of $J^{p}$
values to the above mentioned states are also analyzed. The $J^{p}$
assignment for $D^{0}(2550)$ state is $0^{-}$ while $D^{0}(2600)$ is
identified as mixture of $2^{3}S_{1}$ and $1^{3}D_{3}$ state with
$J^{p}=1^{-}$. The states $D(2750)$ and $D(2760)$ are identified
with $J^{P}=(2^{-},3^{-})$ assignment. All assignments to
$D_{J}(3000)$ are analyzed deeply and various possibilities for
$J^{P}$ states has been checked. Two more possibilities i.e.
$(1^{-},2^{-})$ and $(2^{-},3^{-})$ lying in 1D spectrum have also
been included for analysis. Most possible assignment in the present
work is favored in 2P$(1^{+})$ for $D_{J}(3000)$ state. While
investigating for decays, it is concluded that the results on decay
widths are further helpful to search for un-known resonances so that
the excited meson spectrum for D meson family is clear to theorists
as well as experimentalists.
\section{Acknowledgement}
Part of the work is done under UGC Major Research project NO.SR/
41-959/2012.

\pagebreak

\end{document}